\documentclass{article}
\usepackage[dvips]{graphics}

\newcommand{\bea}{\begin{eqnarray*}}
\newcommand{\eea}{\end{eqnarray*}}
\newcommand{\bean}{\begin{eqnarray}}
\newcommand{\eean}{\end{eqnarray}}
\newcommand{\eqs}[1]{Eqs. (\ref{#1})}
\newcommand{\eq}[1]{Eq. (\ref{#1})}
\newcommand{\meq}[1]{(\ref{#1})}
\newcommand{\fig}[1]{Fig. \ref{#1}}
\newcommand{\ppa}[2]{\left(\frac{\partial}{\partial #1}\right)^{#2}}
\newcommand{\pp}[2]{\frac{\partial #1}{\partial #2}}

\newcommand{\eqn}{&=&}
\newcommand{\non}{\nonumber \\}
\newcommand{\oh}{\frac{1}{2}}

\newcommand{\hsp}{\hspace{0.1mm}}

\newcommand{\tr}{\tilde\rho}

\title{Thin shell model revisited}
\author{Sijie Gao\thanks{ sijie@bnu.edu.cn} and Xiaobao Wang\thanks{1084162983@qq.com} \\
Department of Physics, Beijing Normal University,\\
Beijing 100875, China}
\begin{document}
\maketitle

\begin{abstract}
We reconsider some fundamental problems of the thin shell model. First, we point out that the ``cut and paste'' construction does not guarantee a well-defined manifold because there is no overlap of coordinates across the shell. When one requires that the spacetime metric across the thin shell is continuous, it also provides a way to specify the tangent space and the manifold. Other authors have shown that this specification leads to the conservation laws when shells collide. On the other hand, the well-known areal radius $r$ seems to be a perfect coordinate covering all regions of a spherically symmetric spacetime. However, we show by simple but rigorous arguments that $r$ fails to be a coordinate covering a neighborhood of the thin shell if the metric across the shell is continuous. When two spherical shells collide and merge into one, we show that it is possible that $r$ remains to be a good coordinate and the conservation laws hold. To make this happen, different spacetime regions divided by the shells must be glued in a specific way such that some constraints are satisfied. We compare our new construction with the old one by solving constraints numerically.

\end{abstract}

\section{Introduction}

Since the pioneering work by Israel \cite{israel-66}, the thin shell model has been extensively studied and has a wide application in gravitational collapse, cosmology, wormhole theory, etc. Such a model is important  because it is an idealization of the real matter distribution and has given many interesting solutions in general relativity and alternative gravity theories.

\begin{figure}[htmb]
\centering \scalebox{0.7} {\includegraphics{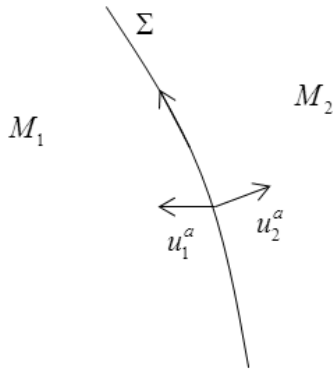}}
\caption{$M_1$ and $M_2$ are connected at $\Sigma$. For each $u_1^a$, one needs to specify a $u_2^a$ such that $u_1^1+u_2^a=0$.} \label{u12}
\end{figure}

Despite great successes of this model,  some fundamental problems still remain to be answered. It is well known that a thin shell is a three dimensional hypersurface  (in four dimension spacetimes) which can be constructed by the `` cut and paste'' approach \cite{israel-91}. Let $M_1$ and $M_2$ be two distinct spacetimes with coordinates $\{x_1^\mu\}$ and $\{x_2^\mu\}$. We can assign a metric $g_{ab}(x_i^\mu)$  to $M_i$, $i=1,2$.  Suppose that each manifold is bounded by a hypersurface $\Sigma_i$. If we wish to unify the two spacetimes, it is natural to glue them by identifying their boundaries, i.e., the new spacetime $M=M_1\cup M_2$ connects the two distinct spacetimes at the hypersurface $\Sigma\equiv\Sigma_1=\Sigma_2$ (see \fig{u12}). There is always a discontinuity of the extrinsic curvature across $\Sigma$ which is related to the surface matter distribution on $\Sigma$. So the first derivative of the metric is discontinuous at $\Sigma$, which leads to the famous junction conditions \cite{israel-66}.  It is also a convention to require that the induce metric $h_{ab}$ is continuous. Howerve, there is no general requirement on the continuity of $g_{ab}$ across $\Sigma$.

Before we go further, we should notice that any metric is defined on a manifold. An overlooked question is: have we had a well-defined manifold by the above construction? If we go through the general properties of manifold, we will find immediately that the answer is ``not yet''.  A manifold allows to be covered by more than one coordinate system. But neither $x_1^\mu$ nor $x_2^\mu$ covers a neighborhood of  a point on $\Sigma$. Although both coordinate systems give coordinates on $\Sigma$, their overlap is only a three dimensional region, not an open set of the manifold \cite{waldbook}. An obvious consequence is that the tangent space of each point on $\Sigma$ is not uniquely defined by the construction so far. Before we glue $M_1$ and $M_2$ together, we only have two ``half tangent spaces'' at their boundaries. Identifying the boundaries does not give a unified tangent space for each point at the boundary. There is an ambiguity for each tangent vector $u_1^a\in V_1 $ at the boundary to find its ``other half'' $u_2^a\in V_2$ satisfying $u_1^a+u_2^a=0$, where $V_1$ and $V_2$ are the tangent spaces of $M_1$ and $M_2$, respectively. There are two equivalent ways to fix the ambiguity. First, we can extend $\{x_1^\mu\}$ to $M_2$ such that there is a four dimensional overlap of $\{x_1^\mu\}$ and $\{x_2^\mu\}$. By this way, $M$ is a well-defined manifold. Second, Note that the three dimensional tangent space of $\Sigma$ has no ambiguity by construction. Thus, we only need to assign one transversal vector $u_1^a$ in $V_1$ with a negative transversal vector $ u_2^a$ in $V_2$ such that $u_1^a+ u_2^a=0$ (see \fig{u12}). Then any other vector is  uniquely assigned a negative vector by the addition rule.

To be definitive, we assume that $\Sigma$ is timelike and the induced metric is given by
\bean
h_{ab}=g^i_{ab}+n^i_an^i_b\,, \ \ \ i=1,2 \,,
\eean
where $n^i_a$ is the spacelike normal of $\Sigma$.
Since $h_{ab}$ is the same from both sides, we see immediately that the spacetime metric $g_{ab}$ is continuous if and only if
\bean
n^1_a=n^2_a \label{n12a}
\eean
By our argument above, \eq{n12a} also uniquely fixes the tangent space of any point on $\Sigma$. In spherically symmetric spacetimes, if a few shells collide, it has been shown by Langlois, Maeda and Wands (LMW) \cite{prl-con} that this identification leads to the conservation of energy and momentum at the collision point. The LMW method has been further applied to bubble and brane collisions \cite{horowitz}-\cite{wanganzhong}.

In a spherically symmetric spacetime, the radial coordinate $r$ is the areal radius of the sphere formed by the SO(3) isometry \cite{waldbook}. So $r$ is a well-defined function and seems to be the only natural coordinate covering different regions divided by the shells. However, we show in section \ref{sec-r}  that if the metric is continuous across the shell, $r$ is no longer a good coordinate for points on the shell. Therefore, if $r$ is a good coordinate, we must choose other identifications which break down the continuity of metrics and generally violate the conservation laws as well. In section \ref{sec-co}, we consider the simplest collision: two shells merge into one after they collide. we derive some constraint equations such that $r$ remains to be a good coordinate and the conservation laws hold. By imposing appropriate initial conditions, we find that these equations are solvable at least numerically.

\section{One spherical thin shell and the $r$ coordinate} \label{sec-r}
\begin{figure}[htmb]
\centering \scalebox{0.7} {\includegraphics{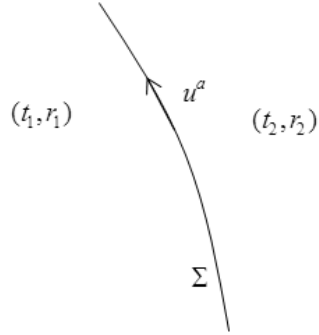}}
\caption{A spherical shell $\Sigma$ moving with four-velocity $u^a$.} \label{fig-tr}
\end{figure}

We consider a spherical shell $\Sigma$ moving in a spherical spacetime. The coordinates on the two sides of the shell are labeled by $(t_1, r_1)$ and $(t_2,r_2)$, where we have dropped the $(\theta, \phi)$ coordinates for simplicity (see \fig{fig-tr}).

The metrics on both sides are in the form
\bean
ds_i^2=-f_i(r)dt^2+f_i^{-1}(r)dr^2+r^2 d\Omega^2
\eean
where $i=1,2$ and in the Schwarzschild case
\bean
f_i(r)=1-\frac{2M_i}{r}
\eean
Note that $r_1=r_2$ by continuity, but $t$ is discontinuous across the shell.  We may write the four-velocity of the shell as
\bean
u^a=\dot t_i \ppa{t_i}{a}+\dot r \ppa{r_i}{a}
\eean
Note that we have used $\dot r$ instead of $\dot r_i$ because $r_1=r_2$. The normalization condition $g_{ab}u^a u^b=-1$ yields
\bean
\dot t_i=\pm \sqrt{\frac{f_i+\dot r^2}{f_i^2}} \label{tdi}
\eean
The normal vector of $\Sigma$ is of the form
\bean
n_i^a=\frac{\dot r}{f_i(r)}\ppa{t_i}{a}+\sqrt{\dot r^2+f_i}\ppa{r_i}{a}
\eean

Now we have three orthogonal and normal tetrads related by the following Lorentz transformation \cite{prl-con}
\bean
 \left(\begin{array}{c}
u^a \\
n^a \end{array} \right)=\Lambda(\alpha_{i})\left(\begin{array}{c}
\sqrt{\frac{1}{f_i}}\ppa{t_i}{a} \\
\sqrt{f_i} \ppa{r_i}{a} \end{array} \right)
\eean
where
\bean
\Lambda(\alpha)=\left(\begin{array}{cc}
\cosh(\alpha) & \sinh(\alpha) \\
\sinh(\alpha) & \cosh(\alpha) \end{array} \right)
\eean
and
\bean
\alpha_i=\sinh^{-1}\frac{\dot r}{\sqrt{f_i}}
\eean
Therefore,
\bean
\left(\begin{array}{c}
\sqrt{\frac{1}{f_2}}\ppa{t_2}{a} \\
\sqrt{f_2} \ppa{r_2}{a} \end{array} \right)=\Lambda(\alpha_2-\alpha_1)\left(\begin{array}{c}
\sqrt{\frac{1}{f_1}}\ppa{t_1}{a} \\
\sqrt{f_1} \ppa{r_1}{a} \end{array} \right) \label{loren}
\eean
We should emphasize that the continuity of metric is crucial to derive this formula. However, we show now that this treatment is inconsistent with the assumption that $r$ is a good coordinate.

As we have mentioned above, the two sets of coordinates $\{t_1,r_1\}$ and $\{t_2,r_2\}$ do not have a four-dimensional overlap in the neighborhood of $\Sigma$. We need first extend  the coordinates smoothly such that they have a four-dimensional overlap region $O_p$ where $p\in \Sigma$. If $r$ is a good coordinate everywhere, we should have
\bean
r_1=r_2=r
\eean
in $O_p$.

Then we can write down the transformation at $p$
\bean
\ppa{t_2}{a}=\pp{t_1}{t_2}\ppa{t_1}{a}+\pp{r_1}{t_2}\ppa{r_1}{a}
\eean
The second term vanishes due to $r_1=r_2$. So
\bean
\ppa{t_2}{a}=\pp{t_1}{t_2}\ppa{t_1}{a} \label{pt2}
\eean

Similarly,
\bean
\ppa{r_2}{a}\eqn \pp{t_1}{r_2}\ppa{t_1}{a}+\pp{r_1}{r_2}\ppa{r_1}{a}\non
\eqn  \pp{t_1}{r_2}\ppa{t_1}{a}+\ppa{r_1}{a} \label{pr2}
\eean
Note that $\alpha_1\neq \alpha_2$ due to $f_1\neq f_2$. Therefore, \eq{loren} indicates that $\ppa{t_1}{a}$ is not parallel to $\ppa{t_2}{a}$. This contradicts \eq{pt2}.

Another quick way to see the breakdown of $r$ is to notice that the $\theta\theta$ component of the extrinsic curvature of $\Sigma$ is given by \cite{gao-lemos}
\bean
K_{\theta\theta}=rn^r=r\sqrt{f+\dot r^2}
\eean
which is obviously discontinuous across the shell. This discontinuity leads to the junction condition. Note that
\bean
n^r=n^a (dr)_a
\eean
So if $n_1^a=n_2^a$, the discontinuity of $n^r$ indicates that $r$ is not a qualified function in any neighborhood of $p\in\Sigma$.

\section{Collision of shells and the conservation laws} \label{sec-co}
In this section, we shall match different sides of shells in a way such that $r$ is a good coordinate across all shells. Then we shall discuss the conservation laws when shells collide. For simplicity, we consider the collision of two shells. After the collision, they merge as one shell.
\begin{figure}[htmb]
\centering \scalebox{0.7} {\includegraphics{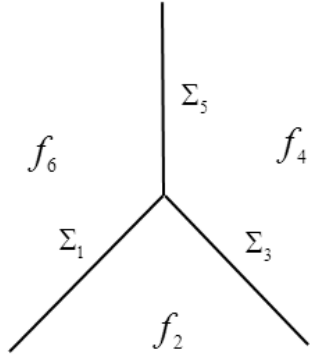}}
\caption{Two shells collide and stick together.} \label{fig-3shell}
\end{figure}

\subsection{Matching conditions}
As shown in \fig{fig-3shell}, $\Sigma_1$ and $\Sigma_3$ represent two shells before the collision and $\Sigma_5$ represents the shell after the collision. The spacetime is then divided into three parts covered by coordinates $\{t_i,r_i\}$, $i=2,4,6$. Applying \eqs{pt2} and \meq{pr2} to $\Sigma_1$, we have
\bean
\ppa{t_2}{a}\eqn T_1\ppa{t_6}{a} \label{d1} \\
\ppa{r_2}{a}\eqn R_1\ppa{t_6}{a}+\ppa{r_6}{a}  \label{d2}
\eean
where $T_1=\pp{t_6}{t_2}$ and $R_1=\pp{t_6}{r_2}$.

Similarly, on $\Sigma_3$ and $\Sigma_5$ we have
\bean
\ppa{t_4}{a}\eqn T_3\ppa{t_2}{a} \label{d3}\\
\ppa{r_4}{a}\eqn R_3\ppa{t_2}{a}+\ppa{r_2}{a} \label{d4} \\
\ppa{t_6}{a}\eqn T_5\ppa{t_4}{a} \label{d5} \\
\ppa{r_6}{a}\eqn R_5\ppa{t_4}{a}+\ppa{r_4}{a} \label{d6}
\eean
Combining \eqs{d5}, \meq{d3} and \meq{d1}, we have
\bean
\ppa{t_6}{a}=T_5T_3T_1 \ppa{t_6}{a}
\eean
while \eqs{d6}, \meq{d4} and \meq{d2} give
\bean
\ppa{r_6}{a}=R_5 T_3T_1\ppa{t_6}{a}+R_3T_1\ppa{t_6}{a}+R_1\ppa{t_6}{a}+\ppa{r_6}{a}
\eean
Therefore,
\bean
T_5T_3T_1\eqn 1  \label{bt1} \\
R_5T_3T_1+R_3T_1+R_1\eqn 0 \label{bt2}
\eean

On the other hand, the four-velocity of $\Sigma_1$ reads
\bean
u_1^a=\dot t_{16} \ppa{t_6}{a}+\dot r_1 \ppa{r_6}{a}=\dot t_{12} \ppa{t_2}{a}+\dot r_1 \ppa{r_2}{a}  \label{u1a}
\eean
where
\bean
\dot t_{16}=\sqrt{\frac{f_6+\dot r_1^2}{f_6^2}},\ \ \  \dot t_{12}=\sqrt{\frac{f_2+\dot r_1^2}{f_2^2}}
\eean
Then junction condition reads
\bean
m_1=r_1\left(\sqrt{f_6+\dot r_1^2}-\sqrt{f_2+\dot r_1^2}\right)
\eean
Substituting \eqs{d1} and \meq{d2} into \eq{u1a} yields
\bean
\dot t_{16} \ppa{t_6}{a}+\dot r_1 \ppa{r_6}{a}=\dot t_{12}T_1\ppa{t_6}{a}+\dot r_1R_1\ppa{t_6}{a}+\dot r_1 \ppa{r_6}{a}
\eean
which gives
\bean
\dot t_{16}=\dot t_{12} T_1+\dot r_1 R_1 \label{td1}
\eean
Similarly, On $\Sigma_3$ and $\Sigma_5$, we have
\bean
\dot t_{32}\eqn \dot t_{34} T_3+\dot r_3 R_3 \label{td2} \\
\dot t_{54}\eqn \dot t_{56} T_5+\dot r_5 R_5  \label{td3}
\eean

\subsection{Conservation of energy and momentum}
The conservation law is
\bean
m_1u_1^a+m_3u_3^a-m_5u_5^a=0 \label{mua}
\eean
Note that
\bean
u_1^a=\dot t_{12} \ppa{t_2}{a}+\dot r_1 \ppa{r_2}{a}
\eean
\bean
u_3^a=\dot t_{32} \ppa{t_2}{a}+\dot r_3 \ppa{r_2}{a}
\eean
\bean
u_5^a\eqn \dot t_{54} \ppa{t_4}{a}+\dot r_5 \ppa{r_4}{a}  \non
 \eqn \dot t_{54} T_3\ppa{t_2}{a}+\dot r_5 \left[R_3\ppa{t_2}{a}+\ppa{r_2}{a}   \right]\non
 \eqn (\dot t_{54} T_3+\dot r_5 R_3)  \ppa{t_2}{a}+\dot r_5 \ppa{r_2}{a}
\eean
where \eqs{d3} and \meq{d4} have been used.

So the conservation law yields
\bean
m_1\dot r_1+m_3\dot r_3-m_5\dot r_5\eqn 0 \label{con1} \\
m_1 \dot t_{12}+m_3 \dot t_{32}-m_5 (\dot t_{54} T_3+\dot r_5 R_3)=m_1 \dot t_{12}+m_3 \dot t_{32}-m_5 \dot t_{52} \eqn 0  \label{con2}
\eean
where we have used
\bean
\dot t_{52}\equiv \pp{t_2}{\tau}\big|_5=\pp{t_2}{t_4}\pp{t_4}{\tau}+\pp{t_2}{r_4}\dot r_5=T_3 \dot t_{54}+R_3 \dot r_5
\eean
in the last step.

\eq{con1} is the $r$ component of \eq{mua}, which is particularly simple. The corresponding equations in the LMW method are not equivalent to our \eqs{con1} and \meq{con2}, although the vector form \meq{mua} of the conservation law is the same in both methods.

\subsection{Solving equations}
Now we have  15 independent variables:
\bea
&&m_1,\dot r_1, m_3,\dot r_3, m_5,\dot r_5 \\
&& f_2, f_4,f_6 \\
&& T_1,T_3,T_5,R_1,R_3,R_5
\eea
while there are 10 equations: 3 junction conditions,  \eqs{bt1} and \meq{bt2}, \eqs{td1},\meq{td2},\meq{td3}, plus two equations of conservation of energy and momentum (\eq{con1} and \meq{con2}).  We may set initial data: $ f_2, f_4,f_6, m_1,m_3$, then the rest variables can be solved.

\subsection{Numerical results}
We take
\bean
f_6=0.7, f_2=0.6, f_4=0.5,\dot r_1=1, \dot r_3=2, r=10. \label{inidata}
\eean
 Then the ten equations mentioned above give rise to the following numerical solutions:
\bean
m_1\eqn 0.3893, \ m_3=0.2344,\ m_5=0.663494, \ \dot r_5=1.2933 \label{m51} \\
T_1\eqn 0.9194,\ \ T_3=0.9208, \ \ T_5=1.1811 \non
R_1\eqn -0.0758, \ R_3=-0.1661, \ R_5=0.2699
\eean

Note that the solutions of $T_i$ and $R_i$ tell us how the manifold is constructed. As we have discussed, this match of manifold differs from the LMW treatment. However, the initial conditions of \eq{inidata} are exactly needed for the LMW method (see Appendix A). It is not surprising that the two methods give rise to different solutions for $m_5$ and $\dot r_5$ (see \eq{m51} and \eq{m52}) because the matching conditions are different.

\section{Conclusions}
In this paper, some fundamental problems of the thin shell model have been reconsidered and clarified. To make the thin shell spacetime a well-defined manifold, some extra conditions need to be imposed. Some authors have proven that the continuity of the metric across the shell leads to the conservation of energy and momentum in spherically symmetric spacetimes. However, we show that in this treatment, the areal radius $r$ is no longer a coordinate covering a neighborhood of the shell. We have then proposed a new matching technique such that the conservation law and the coordinate $r$ are              both preserved. In the case that two shells collide and merge into one shell, we have shown that the initial conditions that needed to solve all the equations are exactly the same as in the LMW method. Our work suggests that spacetimes containing thin shells can be matched in different ways and the conservation laws can still be preserved.

\section*{Acknowledgements}
This research was supported by NSFC Grants No. 11235003, 11375026 and NCET-12-0054.

\appendix
\section{ Review of LMW mechanism and numerical calculation}
In this appendix, we review LMW's treatment and apply it to the case in section \ref{sec-co}. With the same initial conditions, the numerical computation shows that the two methods gives different results.

\subsection{One shell}
The junction condition of one shell is
\bean
\sqrt{f_1+\dot r^2}-\sqrt{f_2+\dot r^2}=\rho\hsp r
\eean
where $r$ increases from region $1$ to region $2$ and $\rho$ is the surface       density of the shell.
Let
\bean
\sinh\alpha_i=\frac{\dot r}{\sqrt{f_i}} \label{sinh}
\eean
we have
\bean
\sqrt{f_i+\dot r^2}=\sqrt{f_i}\sqrt{1+\sinh^2\alpha_i}=\sqrt{f_i}\cosh\alpha_i     \eean
Let
\bean
\tr=\rho r
\eean
Then
\bean
\sqrt{f_1}\cosh\alpha_1-\sqrt{f_2}\cosh\alpha_2=\tr
\eean
i.e.,
\bean
\oh \sqrt{f_1}e^{\alpha_1}+\oh \sqrt{f_1}e^{-\alpha_1}-\oh \sqrt{f_2}e^{\alpha_2}-\oh \sqrt{f_2}e^{-\alpha_2}\eqn \tr  \label{rho12}
\eean
\eq{sinh} leads to
\bean
\frac{e^{\alpha_1}-e^{-\alpha_1}}{e^{\alpha_2}-e^{-\alpha_2}}=
\frac{\sqrt{f_2}}{\sqrt{f_1}}
\eean
So \eq{rho12} may be written as
\bean
 \sqrt{f_1}e^{\alpha_1}- \sqrt{f_2}e^{\alpha_2}\eqn \tr \label{tra} \\
  \sqrt{f_1}e^{-\alpha_1}- \sqrt{f_2}e^{-\alpha_2}\eqn \tr
\eean

\subsection{Three shells}
We still consider the collision of two shells as shown in \fig{fig-3shell}. Applying \eq{sinh} to each shell, we have
\bean
\sinh\alpha_{16}\eqn \frac{\dot r_1}{\sqrt{f_6}}, \ \ \ \sinh\alpha_{12}= \frac{\dot r_1}{\sqrt{f_2}} \label{sinh1} \\
\sinh\alpha_{32}\eqn \frac{\dot r_3}{\sqrt{f_2}}, \ \ \ \sinh\alpha_{34}= \frac{\dot r_3}{\sqrt{f_4}}\label{sinh2} \\
\sinh\alpha_{54}\eqn \frac{\dot r_5}{\sqrt{f_4}}, \ \ \ \sinh\alpha_{56}= \frac{\dot r_5}{\sqrt{f_6}} \label{sinh3}
\eean

Applying \eq{tra} to each shell yields
\bean
\tr_1\eqn \sqrt{f_6} e^{\alpha_{16}}-\sqrt{f_2} e^{\alpha_{12}} \\
\tr_3\eqn \sqrt{f_2} e^{\alpha_{32}}-\sqrt{f_4} e^{\alpha_{34}} \\
\tr_5\eqn \sqrt{f_4} e^{\alpha_{54}}-\sqrt{f_6} e^{\alpha_{56}}
\eean
Note that $\tr_5$ given above has a sign difference from what we gave previously because it corresponds to the shell after the collision.

The consistency condition is given by \cite{prl-con}
\bean
\alpha_{16}-\alpha_{12}+\alpha_{32}-\alpha_{34}+\alpha_{54}-\alpha_{56}=0 \label{azero}
\eean
If we define
\bean
\alpha_{ij}=-\alpha_{ji}
\eean
The condition becomes
\bean
\alpha_{12}+\alpha_{23}+\alpha_{34}+\alpha_{45}+\alpha_{56}+\alpha_{61}=0 \label{con6}
\eean

For $f_2$,
\bean
&&\tr_1 e^{\alpha_{23}}e^{\alpha_{34}}e^{\alpha_{45}}e^{\alpha_{56}}e^{\alpha_{61}}+\tr_3 e^{\alpha_{23}}+\tr_5 e^{\alpha_{23}}e^{\alpha_{34}}e^{\alpha_{45}} \non
\eqn \sqrt{f_6} e^{\alpha_{16}} e^{\alpha_{23}}e^{\alpha_{34}}e^{\alpha_{45}}e^{\alpha_{56}}e^{\alpha_{61}}-\sqrt{f_6} e^{\alpha_{56}}e^{\alpha_{23}}e^{\alpha_{34}}e^{\alpha_{45}}+...=0
\eean
where we have used \eq{con6}. Such relations can be written as
\bean
\tr_1 e^{\pm\alpha_{21}}+\tr_3 e^{\pm\alpha_{23}}+\tr_5 e^{\pm\alpha_{25}}=0 \label{r135}
\eean
where $\alpha_{25}=\alpha_{23}+\alpha_{34}+\alpha_{45}$.
The other two relations can be obtained by replacing $2$ by $4$ and $6$.  It is also easy to get
\bean
\tr_1\gamma_{21}+\tr_3\gamma_{23}+\tr_5\gamma_{25}=0
\eean
where $\gamma_{ij}=\cosh\alpha_{ij}$. This is the energy conservation law
\cite{prl-con}.

\subsection{Numerical results}
We use the same initial data as given in \eq{inidata}. Then \eqs{sinh1} and \meq{sinh2} can be solved directly. Combining \eq{sinh3} and \eq{azero}, one can obtain $\alpha_{54}=1.3876$ and $\alpha_{56}=1.2421$ . Then we have
\bean
m_5=0.6506, \ \ \ \dot r_5=1.3278 \label{m52}
\eean
which is different from the previous results.


\begin{thebibliography}{99}
\bibitem{israel-66} W.Israel, Nuovo Cimento B {\bf 44,} 1 (1966).
\bibitem{israel-91} C.Barrabes, W.Israel, Phys. Rev. D, {\bf 43,} 1129 (1991).
\bibitem{waldbook} R.M. Wald,{\em General Relativity} (The University of Chicago Press, Chicago, 1984).
\bibitem{prl-con} D.Langlois, K.Maeda and D.Wands, Phys. Rev. Lett. {\bf 88,} 181301 (2002).
\bibitem{gao-lemos} S.Gao and J.P.S.Lemos, IJMPA, {\bf 23,} 2943(2008).
\bibitem{horowitz} B.Freivogel, G.T.Horowitz and S.Shenker, JHEP 05(2007)090.
\bibitem{tk} Takamizu, K.Maeda, Phys.Rev.  D {\bf 70,} 123514(2004).
\bibitem{wanganzhong} A.Tziolas and A.Wang, Phys. Lett. B,{\bf  661,} 5-10 (2008).
\end{thebibliography}
\end{document}